\documentclass{PoS}

\usepackage{amsmath}

\title{Composite Higgs theory}

\ShortTitle{Composite Higgs theory}

\author{\speaker{Florian Goertz}
\\
        Max-Planck-Institut f{\"u}r Kernphysik, Saupfercheckweg 1, 69117 Heidelberg, Germany\\
        E-mail: \email{fgoertz@mpi-hd.mpg.de}}

\abstract{I review the idea of realizing the Higgs
as a composite pseudo-Nambu-Goldstone boson of a new 
strongly-interacting sector and collect major constraints on the
parameter space of minimal models. Besides limits from electroweak
precision tests, LHC searches for resonances and bounds due to 
Higgs-coupling modifications will be discussed in detail. 
Finally, the issue of light top partners in these models will be
explored, including ways to avoid them which lead to interesting
implications for flavor observables.}

\FullConference{An Alpine LHC Physics Summit (ALPS2018)\\
		15-20 April, 2018\\
		Obergurgl, Austria}

\begin{document}

\section{Introduction}
To understand the stability of the weak scale, $v=(\sqrt 2
G_F)^{-1/2} \approx 246\,$GeV, given the presence of large
thresholds, such as the Planck scale $M_{\rm Pl}\sim 
10^{19}$\,GeV or the scale of grand unification $M_{\rm GUT}
\sim 10^{15}$\,GeV, remains a major task in high energy physics.
Beyond this notorious `hierarchy problem', there are further
hierarchies in the flavor sector, in particular in quark masses
and mixings and due to the tiny neutrino masses, that can not be
explained in the Standard Model (SM) of particle physics.
Moreover, in the SM, electroweak symmetry breaking (EWSB) -- being
responsible for the masses of the known elementary particles -- 
is just parameterized via the Higgs mechanism, but not explained
dynamically.

All these issues can be addressed in models where EWSB is not
triggered by a fundamental elementary scalar, but ultimately
induced by the condensate of a new strong interaction 
$\langle \hat O \rangle$, breaking spontaneously a global symmetry 
\begin{equation}
G \xrightarrow{\ \langle \hat O \rangle \ } H
\end{equation}
with the electroweak group $G_{\rm EW} \equiv SU(2)_L \times
U(1)_Y$ embedded in $G$ as a weakly gauged subgroup. The prime
example of such a mechanism of dynamical EWSB is realized in
Technicolor (TC) theories \cite{Weinberg:1975gm,Susskind:1978ms}, 
which furnish a good starting point to discuss more recent 
incarnations of EWSB via similar kinds of dynamics. In the former
theories, $G_{\rm EW}$ is fully broken along with $G$ at the scale 
$f \sim v $ via condensation, by upscaling the breaking pattern 
\begin{equation}
G=SU(2)_L \times SU(2)_R \, \xrightarrow{\langle \bar q q \rangle}
 \, H = SU(2)_V
\end{equation}
of QCD to the EW scale. The latter is now generated from
a UV theory via dimensional transmutation, i.e., emerging from a
new running coupling becoming strong at a certain scale and triggering the condensation
of new (EW charged) 'techni'-fermions, $\langle \bar q_{\rm TC}
\,q_{\rm TC} \rangle \neq 0$, which breaks EW symmetry.
In this way three Goldstone bosons ('techni-pions'), delivering
the longitudinal degrees of freedom for the massive $W$ and $Z$
bosons, emerge by breaking three $SU(2)$ generators
(belonging to the coset $G/H$). The large Goldstone
decay constant $f\simeq v=246$\,GeV $\gg f_\pi = 92$\,MeV finally allows
for viable weak boson masses.
Yet, no physical Higgs boson emerges to explain the 2012
discovery at the LHC \cite{Chatrchyan:2012xdj,Aad:2012tfa} (see,
however \cite{Holdom:1981rm,Bardeen:1985sm,Yamawaki:1985zg,Bellazzini:2012vz}).
Moreover, TC theories suffer generically from large corrections to
EW precision parameters and it is a challenge to include fermion masses.

All these problems can be solved in {\it composite Higgs} (CH)
models \cite{Kaplan:1983fs,Kaplan:1983sm,Dugan:1984hq}, which
employ a larger coset $G/H$ of broken global symmetries (like
$SO(5)/SO(4)$) such as to deliver at least four Goldstone bosons
that can furnish a full (composite) Higgs doublet. In turn,
EW symmetry $G_{\rm EW} \subset H$ is now broken via the
vacuum misalignment mechanism, as detailed in the next section.
The Higgs potential emerges radiatively via explicit breaking of
the Goldstone symmetry, making the Higgs a pseudo Nambu-Goldstone
boson (pNGB). The hierarchy problem is still solved since the
Higgs boson is not a fundamental scalar, but is composite above
the TeV scale and its mass
is thus saturated in the IR. Moreover, its Goldstone nature 
provides a reasoning for its lightness compared to other new
states. Finally, the presence of a Higgs doublet and the related
possibility to separate the EW scale $v$ from the global
symmetry breaking scale $f$ allows for a suppression of
corrections to (precision) observables by the ratio $\xi \equiv
v^2/f^2$ and eventually for a SM limit, decoupling the heavy
resonances that reside at $m_\ast \sim g_\ast f$, with
$g_\ast < 4 \pi$ the coupling of the composite sector (at the
price of reintroducing fine-tuning).

The remainder of these notes is organized as follows. In 
Section~\ref{sec:CH} we go on to introduce the CH idea and in
particular the vacuum misalignment mechanism and the resulting
Higgs potential in more detail. We will also discuss corrections
to the Higgs couplings due to its Goldstone nature and briefly
introduce the concept of partial compositeness to realize fermion
masses. After that, in Section \ref{sec:EWPT}, we will explore
constraints on $\xi$ following from electroweak precision tests 
(EWPT), while in Section \ref{sec:LHC}, we will present current 
limits from LHC searches for heavy resonances and for modified 
Higgs couplings. CH models generically predict the presence of
anomalously light top partners with masses $m_{t^\prime} < f < 
m_\ast$, which start to be in conflict with null-searches at the
LHC. Section \ref{sec:tp} contains a discussion of a CH
incarnation that avoids such ultra-light partners, while
addressing neutrino masses via a seesaw mechanism, with potentially
interesting consequences for flavor physics. Finally, Section
\ref{sec:conc} contains the conclusions.
Although these notes are meant to be self-contained, the emphasis is 
on providing a condensed overview of the theoretical setup of CH models
and current constraints. For more details on the discussed topics, the 
reader is referred to the cited literature and comprehensive reviews,
such as \cite{Contino:2010rs,Panico:2015jxa}.

\section{Vacuum Misalignment and a Composite Higgs}
\label{sec:CH}

\begin{figure}[!t]
	\begin{center}
	\includegraphics[height=1.8in]{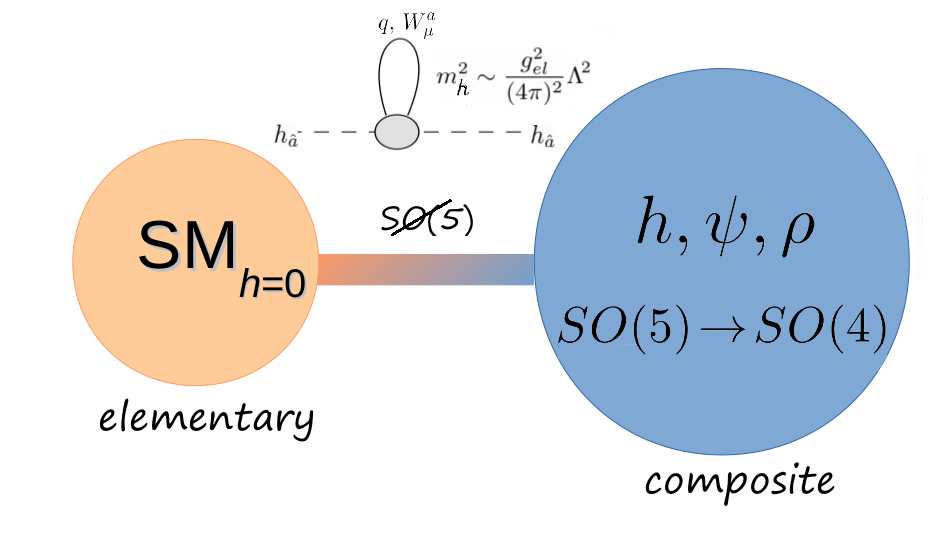} 
	\caption{\label{fig:CHpic} Pictorial representation of the
	Composite Higgs setup, see text for details.}
	\end{center}
\end{figure}

\begin{figure}[!t]
	\begin{center}
	\includegraphics[height=1.85in]{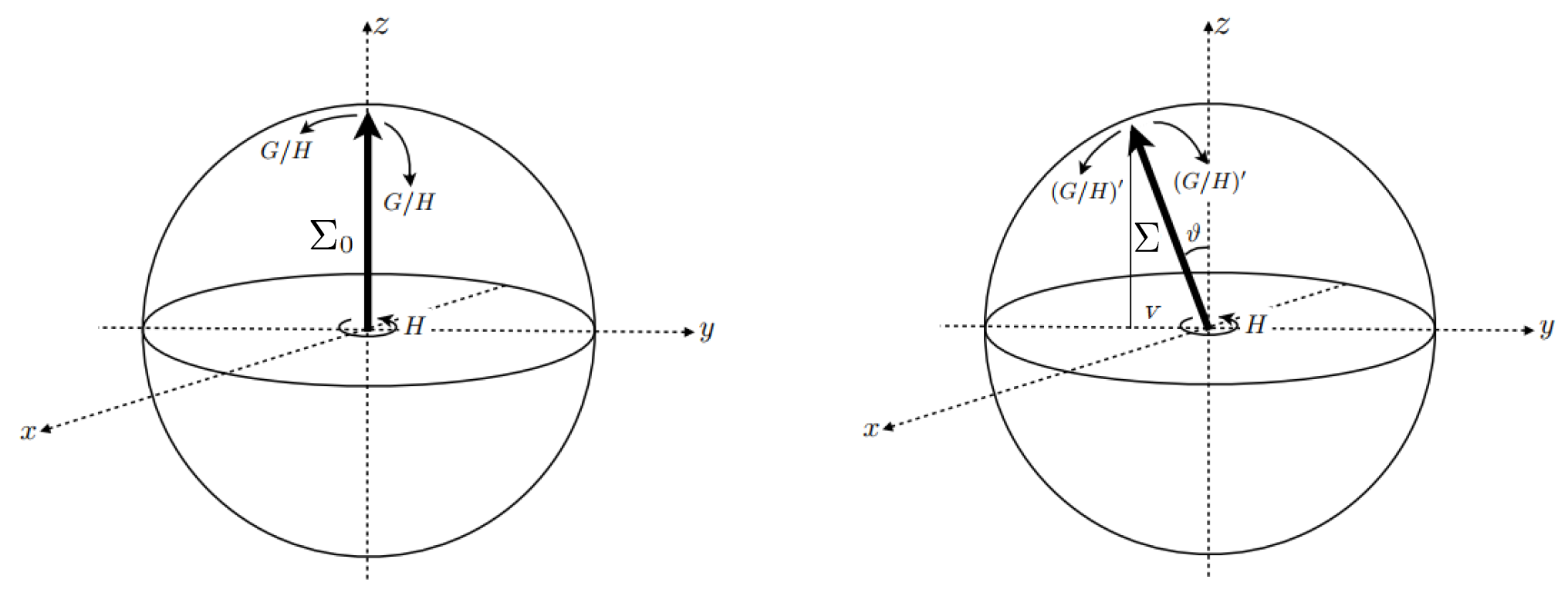} 
	\caption{\label{fig:misalign} Vacuum misalignment, breaking
	$G_{\rm EW} \subset H$. Picture adapted from
	\cite{Azatov:2012qz}, see text for details.}
	\end{center}
\end{figure}

While several non-minimal cosets have been studied in the
literature (see \cite{Bellazzini:2014yua} for an overview), here
we focus on the minimal composite Higgs realization that features
custodial symmetry, {\it i.e.,} $G/H=SO(5)/SO(4)$ 
\cite{Agashe:2004rs}. We thus consider a strongly coupled sector
that induces $SO(5) \to SO(4) (\simeq SU(2)_L \times 
SU(2)_R)$ breaking and contains the composite Higgs as well as 
bosonic and fermionic resonances, $\rho$ and $\psi$, respectively. 
This sector is coupled to the SM, which breaks explicitly the 
global symmetry, since i) the SM gauges only a subgroup of $SO(5)$ 
and ii) the SM fermions don't fill complete $SO(5)$ representations, 
and thus induces radiatively a potential for the Goldstone Higgs, 
which in turn breaks EW symmetry and provides masses for the EW gauge 
bosons and SM-like fermions. The setup is summarized in pictorial 
form in Fig.~\ref{fig:CHpic}.

Explicitly, the {\it minimal} description of the composite pNGB
Higgs corresponds to a non-linear $\sigma$-model of the 
$SO(5)/SO(4)$ coset (see, e.g., \cite{Coleman:1969sm,Callan:1969sn,Panico:2011pw,Giudice:2007fh,Barbieri:2007bh,Carmona:2014iwa}). 
The respective Goldstone bosons are parameterized by the 
$\Sigma$-field
\begin{equation}
\Sigma = U \Sigma_0 \,,
\end{equation} 
which contains the Goldstone matrix
\begin{equation}
U= {\rm exp} \left(i \frac{\sqrt 2}{f} h_{\hat a} 
T^{\hat a}\right) \,,
\end{equation}
and corresponds to a local rotation of the ($SO(4)$-preserving)
vacuum configuration
\begin{equation}
\Sigma_0 = (0,0,0,0,f)^T\,.
\end{equation}
The Goldstone fields $h_{\hat a}$ are in fact just the 'angular'
variables associated to local transformations in the direction of
the broken $SO(5)$ generators $T^{\hat a}_{ij} \equiv -i/\sqrt 2
\left[\delta_i^{\hat a} \delta_j^5 - \delta_j^{\hat a}
\delta_i^5\right]$, and feature the correct quantum numbers to
furnish a Higgs doublet.
The corresponding Lagrangian, replacing the Higgs sector in the 
SM, reads
\begin{equation}
{\cal L}_\Sigma=\frac 1 2 \left(D_\mu \Sigma\right)^T 
D^\mu \Sigma\,,
\end{equation}
where $D_\mu=\partial_\mu -i g^\prime\, Y B_\mu -i g\, T^i 
W^i_\mu$ is the gauge-covariant derivative, which induces the
couplings of the composite Higgs to the SM gauge fields.

Although this minimal description is not complete since it does
for example not include the composite resonances $\rho,\Psi$ of
the new strongly coupled sector, it allows to derive already 
several characteristic predictions of the CH scenario, as we 
will see below. More UV complete models, including this next
threshold (i.e., the heavy resonances) have been constructed 
in the literature, such as the 5D holographic duals 
\cite{Contino:2003ve,Agashe:2004rs} or the (deconstructed) 
2-site / 3-site models \cite{Panico:2011pw} (see also
\cite{DeCurtis:2011yx}), and the reader is referred to the
corresponding articles for more details. 
Still, we will at least also introduce a layer of fermionic
resonances later when we discuss the realization of the fermion
sector via partial compositeness. 

Before deriving first predictions of the scenario let us review 
in more detail how EW is finally broken in 
CH models, following the geometrical picture in 
Fig.~\ref{fig:misalign}. The spontaneous breaking of $G$ triggered
by strong dynamics leaves -- without explicit $G$
breaking -- a global symmetry $H$ unbroken, with the vacuum 
$\Sigma_0$ pointing in a direction orthogonal to $H$ (see left
panel) and the Goldstone bosons of the $G/H$ coset being exactly
massless. However, due to explicit $G$-breaking via gauging of
$G_{\rm EW} \subset H \subset G$ and couplings to the SM fermions
(see below), the Goldstone-Higgs develops a potential and a vacuum
expectation value (vev) $\langle h_{\hat a}^2 \rangle >0$, 
breaking $G_{\rm EW} \subset H$ by shifting the true vacuum 
$\langle \Sigma \rangle$ with respect to the $H$-preserving 
$\Sigma_0$ by an angle 
\begin{equation}
\vartheta \equiv \langle h\rangle/f\,,
\end{equation}
where $h \equiv \sqrt{(h_{\hat a})^2}$.
This is visualized in the right panel of Fig.~\ref{fig:misalign}.
The amount of breaking of EW symmetry now corresponds to the 
projection of the shifted vacuum onto the $H$-plane
\begin{equation}
v = f \sin \vartheta\,,\quad f = |\Sigma_0|\,,
\end{equation}
with $v \approx 246\,$ GeV the EW vev. 
The latter measures the misalignment of the true vacuum with 
respect to $\Sigma_0$ and the corresponding mechanism is referred 
to as \emph{vacuum misalignment mechanism}~\cite{Kaplan:1983fs,Kaplan:1983sm,Dugan:1984hq}.
As we sill see below, the challenge now becomes to generate a
small value
\begin{equation}
\xi = \frac{v^2}{f^2} = \sin^2 \vartheta \ll 1\,,
\end{equation}
such as to abandon the TC limit $\xi \to 1$ and suppress
corrections to SM predictions, scaling with $\xi$.

\paragraph{Higgs couplings}
In fact, we are now ready to have a first look on the couplings of
the pNGB Higgs to the SM-like gauge bosons.
At low energies we can conveniently describe the properties of the
latter (in the background of $\Sigma$) using
symmetries \cite{Agashe:2004rs,Contino:2010rs}. 
Promoting the full global $SO(5) \times U(1)_X$ of the strong 
sector to a gauge symmetry\footnote{The additional $U(1)_X$ factor
is needed to allow for viable hypercharges of the SM fermions and 
we will later turn of the spurious gauge degrees of freedom.}, the
most general (quadratic) Lagrangian takes the form
\begin{equation}
\label{eq:FF}
{\cal L}_{\rm eff}^V = \frac{1}{2} (P_T)^{\mu\nu} \left[ 
\Pi_0^X(q^2)X_\mu X_\nu +
\Pi_0(q^2){\rm Tr}(A_\mu A_\nu) + \Pi_1(q^2)\Sigma A_\mu A_\nu \Sigma^T
\right]
 \,,
\end{equation}
with $X$ and $A_\mu$ the $U(1)_X$ and $SO(5)$ gauge bosons,
respectively, $(P_T)^{\mu\nu} \equiv \eta^{\mu\nu}- 
q^\mu q^\nu/q^2$ , and $\Sigma$ is treated as a classical 
background (with vanishing momentum). 

Using symmetries and results
valid for large number of 'colors' $N$ in the strong sector
\cite{tHooft:1973alw,tHooft:1974pnl,Witten:1979kh}, 
one can derive explicit results for the form factors $\Pi(q^2)$,
that encode the strong dynamics, expanding them for low momenta.
In particular, one finds $\Pi_1(0)=f^2$ and, using the properties 
of the $SO(5)/SO(4)$ generators $T^{\hat a}$ as well as switching 
of the unphysical gauge fields, we finally obtain, following 
\cite{Contino:2010rs} 
\begin{equation}
\begin{split}
{\cal L}_{\rm eff}^V  = &\ \frac{f^2}{8} \sin^2\left(\frac{\langle h \rangle + h}{f}\right) (W_\mu^i W^{i \mu} - 2 W_\mu^3 B^\mu + B_\mu B^\mu) + \cdots \ \\[0.4mm]
 = & \ (1+ 2\sqrt{1-\xi}\,\frac{h}{v} + (1-2 \xi)\,\frac{h^2}{v^2} + \cdots) \left(m_W^2 W_\mu^+ W^{-\mu} + \frac{m_Z^2}{2} Z_\mu Z^\mu \right) + \cdots \,,
\end{split}
\end{equation}
where we have absorbed the gauge couplings into the normalization 
of the corresponding fields in the intermediate steps. We observe 
that the couplings of the pNGB Higgs to gauge bosons are generically
reduced compared to the corresponding interactions in the 
SM.\footnote{See \cite{Liu:2018vel} for a discussion on the 
universality of these results in the IR, i.e., their independence of 
the concrete coset.}
For the couplings of one (two) Higgs bosons to EW gauge fields, 
$g_{hVV}$ $(g_{hhVV})$, we obtain
\begin{equation}
\label{eq:HVV}
g_{hVV} = \sqrt{1-\xi}\, g_{hVV}^{\rm SM}  \,,\quad g_{hhVV} = (1-2 \xi)\, g_{hhVV}^{\rm SM}\,,
\end{equation}
which approach the SM values for $\xi \to 0$.

\paragraph{Partial compositeness and the Higgs potential}

We now turn to the implementation of fermions in the CH framework.
It turns out that the framework of \emph{partial compositeness} is
particularly suited to realize fermion masses, delivering even an 
explanation for the large hierarchies that are observed among them
\cite{Kaplan:1991dc,Agashe:2004rs,Contino:2003ve,Contino:2006qr}.
Focusing on the quark sector, one basically assumes that the elementary
SM-like fields $q_{L,R}$ mix linearly with composite resonances 
$\Psi^{Q,q}_{L,R}$, with the corresponding mass-mixing Lagrangian 
reading \cite{Carmona:2014iwa} (see also \cite{Panico:2011pw,Contino:2006qr,Azatov:2011qy})
	\begin{eqnarray}
	\label{eq:Lmass}
	 {\cal L}_{\rm mass}^\Psi & = &
	 -\, y_L^q f\  \bar q_L\, \Delta_L^q \Psi_R^Q\, -\, y_R^q f\ \bar q_R\, \Delta_R^q
	\Psi^q_L  \\[2mm]
	& & - \text{\small $\sum_{f,f^\prime=Q,q}$}  m_\Psi^{f f^\prime}\, \bar \Psi_L^f \Psi_R^{f^\prime}   \
	- f\!\!\text{\small $\sum_{f,f^\prime=Q,q}$}\! Y_{f f^\prime} \	
\bar \Psi_L^f  \frac{\Sigma \Sigma^T}{f^2} \Psi_R^{f^\prime}	
	 \, + {\rm h.c.}\,. \nonumber
	\end{eqnarray}
Here, we assumed the composite resonances to reside in the fundamental
representation {\bf 5} of $SO(5)$ -- otherwise the Yukawa couplings 
in the strong sector $\sim Y_{f f^\prime}$ have a different form -- 
and the objects $\Delta_{L,R}^q$ project out the components
of $\Psi^{Q,q}$ that can couple to the SM fermions in a $G_{\rm EW}$
invariant way. Note that, in the literature, the minimal (i.e., $SO(5)/SO(4)$)
composite Higgs model with fermions in the {\bf 5} is denoted as 
MCHM$_5$.

\begin{figure}[!t]
	\begin{center}
	\includegraphics[height=1.1in]{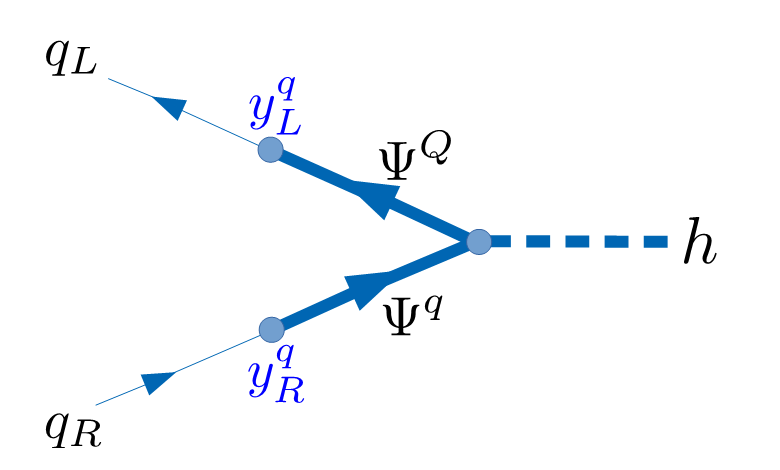}\qquad 
	\includegraphics[height=1in]{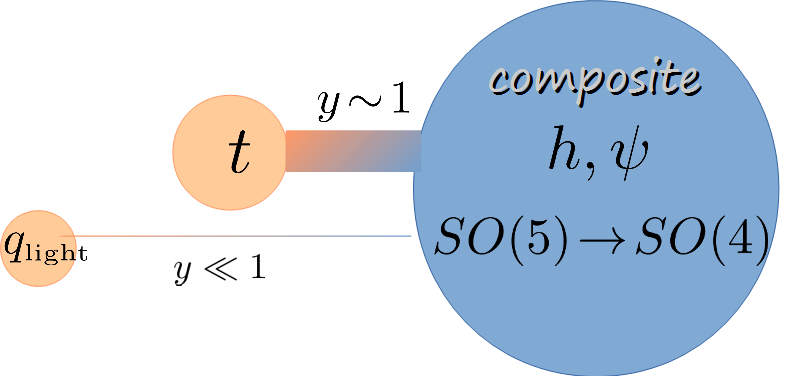} 
	\caption{\label{fig:PC} Generation of fermion masses via
	partial compositeness, see text for details.}
	\end{center}
\end{figure}

From Eq.~\eqref{eq:Lmass} follows that, after diagonalization, the 
SM-like quarks now correspond to a mixture of elementary and composite
fields and the quark masses take the form
\begin{equation}
\label{eq:mtop}
m_q \sim y_L^q y_R^q\, Y_{Qq}\, \frac{v f^2}{m_\Psi^{QQ} m_\Psi^{qq}}\,,
\end{equation}
which can also be obtained from integrating out the heavy resonances
in the diagram in the left panel of Fig.~\ref{fig:PC}.
Now, hierarchically different degrees of compositeness $y^q_{L,R}$ of 
the quarks lead to hierarchical mass eigenvalues (and mixings),
see the right panel of the figure, and the
former in fact arise in strongly coupled theories via renormalization group running from {\it small}
differences in anomalous dimensions of the associated composite 
operators  \cite{Agashe:2004rs}. 
In this way, CH models address -- besides the gauge hierarchy 
problem -- also the flavor puzzle.\footnote{In the dual 5D theory 
the varying degree of compositeness corresponds to different 
fermionic wave-function localizations in the extra dimension,
originating from ${ \cal O}(1)$ input parameters (see, e.g., \cite{Gherghetta:2010cj} and references therein), and the structure
closely resembles that of the Froggatt-Nielsen mechanism 
\cite{Grossman:1999ra,Huber:2000ie,Csaki:2008zd,Casagrande:2008hr}.}

The $\Delta_{L,R}^q$ in Eq.~\eqref{eq:Lmass} explicitly break the 
$SO(5)$ Goldstone symmetry and thus generate a potential for the pNGB
Higgs, which is in general a combination of trigonometric functions of
$h/f$. It turns out that the coefficients of these functions need to
cancel to (at least) the order $\sin^2(v/f) \ll 1$ to allow for 
$0<v<<f$, because in general the vacuum, once shifted, tends to 
be maximally misaligned, $v \sim f$,  with respect to the
the $SO(4)$ preserving one, due to the explicit symmetry breaking
\cite{Agashe:2004rs,Panico:2011pw,Carmona:2014iwa}. Clearly, the most
important contribution comes from the field with the largest
compositeness, which is the heavy top quark, and here and in the
following we will neglect the subleading contributions from lighter
quarks (and gauge bosons). 

In the end, also the Higgs mass will be
proportional to the Goldstone symmetry breaking ($\sim y_{L,R}^t$), 
and after a careful evaluation one obtains for the MCHM$_5$
\cite{Panico:2011pw,Matsedonskyi:2012ym,Carmona:2014iwa}
\begin{equation}
m_h \sim y_t^2 v \sim \frac{m^0_T}{f} m_t\,,
\end{equation}
where we employed Eq.~\eqref{eq:mtop} and used the fact that 
$y_L^t \sim y_R^t \equiv y_t$ and that $Y_{Qq}$ can be expressed 
in terms of resonance masses \cite{Matsedonskyi:2012ym}, while 
$m^0_T \equiv {\rm min}(m_\Psi)$ is the mass of the 
lightest 'top-partner' resonance. Since one expects the latter to 
reside in general above the scale $f$, we observe that generically 
the Higgs boson is expected to be heavier than the top quark. A
phenomenologically viable $m_h \sim 125\,$GeV requires -- at 
least in the simplest CH realizations -- the lightest top-sector
resonance to feature a mass $m^0_T < f$ \cite{Contino:2006qr,Matsedonskyi:2012ym,Pomarol:2012qf,Csaki:2008zd,DeCurtis:2011yx}. 
This is thus in particular much lighter than the vector resonances
but allows in turn for a viable top mass with a reduced Goldstone symmetry
breaking, as needed for a light Higgs. 
For $f \sim \,$TeV, the above estimate is 
however in tension with LHC null
results in resonance searches and, as we will detail below, it is
interesting to search for models which avoid the presence of such
problematically {\it light top partners}.
Before exploring in more detail the phenomenology of the latter,
we will first collect lower bounds on $f$ from various observables,
such as to understand if CH models can 
in general (still) be realized at the
1~TeV scale.

\section{Electroweak Precision Tests}
\label{sec:EWPT}

CH models can be tested in precise extractions of electroweak 
(pseudo-)observables, such as the $S,T,U$ parameters, which parameterize
NP contributions to electroweak vacuum polarization diagrams
\cite{Peskin:1990zt,Peskin:1991sw,Haller:2018nnx}.
Due to custodial symmetry, the tree-level contribution to $T$ vanishes 
in the MCHM, while the tree-exchange of spin-1 resonances 
generates a positive \cite{Contino:2010rs,Agashe:2004rs}
\begin{equation}
S = 2 \pi \xi\, \Pi_1^\prime(0) \approx 4 \pi \frac{v^2}{m_\rho^2} \,,
\end{equation} 
which depends on the form factor introduced in Eq.~\eqref{eq:FF}. Moreover,
$m_\rho \sim m_\ast$ is the scale of the vector-resonances,\footnote{
This does not necessarily coincide with the scale of fermionic
resonances, since the sectors might in principle be governed by 
different couplings, $g_\rho \neq g_\Psi$.} and, as opposed to the 
TC case, $S$ can become arbitrarily small considering~$\xi~\ll~1$.

A second important contribution to the electroweak precision 
parameters arises at one-loop from the fact that the Higgs to 
gauge-boson couplings are modified \cite{Barbieri:2007bh}, 
$g_{VVh}/g_{VVh}^{\rm SM} = \sqrt{1-\xi} \neq 1$. One finds
\cite{Barbieri:2007bh,Contino:2010rs}
\vspace{-3mm}
\begin{equation}
\begin{split}
\Delta S =& + \frac{1}{12 \pi} \xi \log\frac{\Lambda^2}{m_h^2}\,, \\
\Delta T =& - \frac{3}{16 \pi \cos^2 \theta_W} \xi \log\frac{\Lambda^2}{m_h^2}\,,
\end{split}
\end{equation}
with $\Lambda \approx 4 \pi f$.
Recent global analyses of constraints on the MCHM from EWPT arrive at \cite{Thamm:2015zwa,Ghosh:2015wiz}
\begin{equation}
\xi \lesssim 0.1 \ \Leftrightarrow \ f \gtrsim 800\,{\rm GeV} \ \, @ 95\% {\rm CL}\,.
\end{equation}

\vspace{-4mm}

\section{LHC Searches}
\label{sec:LHC}

\vspace{-2mm}
\subsection{Resonance Searches}

\begin{figure}[!t]
	\includegraphics[height=1.3in]{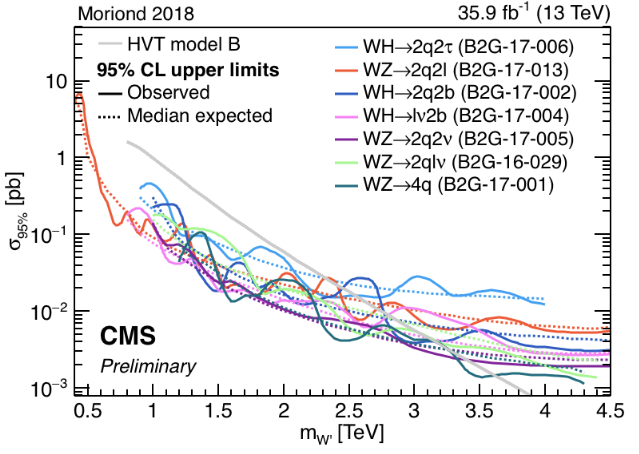}\,
	\includegraphics[height=1.26in]{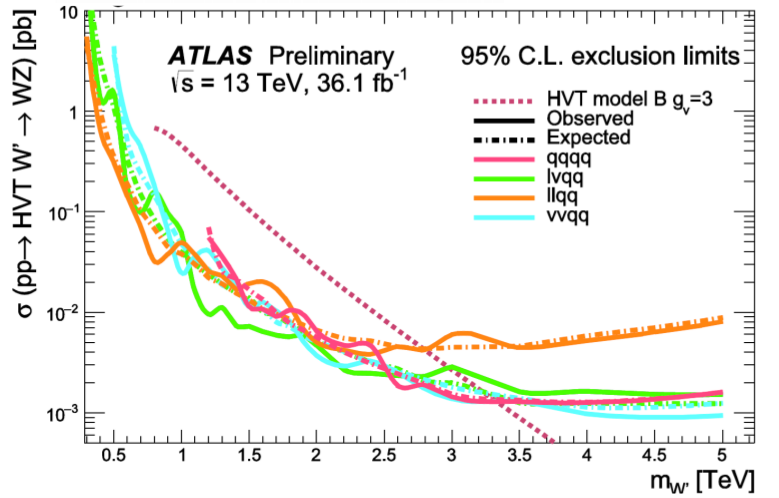}\quad
	\raisebox{-1.3cm}{\includegraphics[height=1.8in]
	{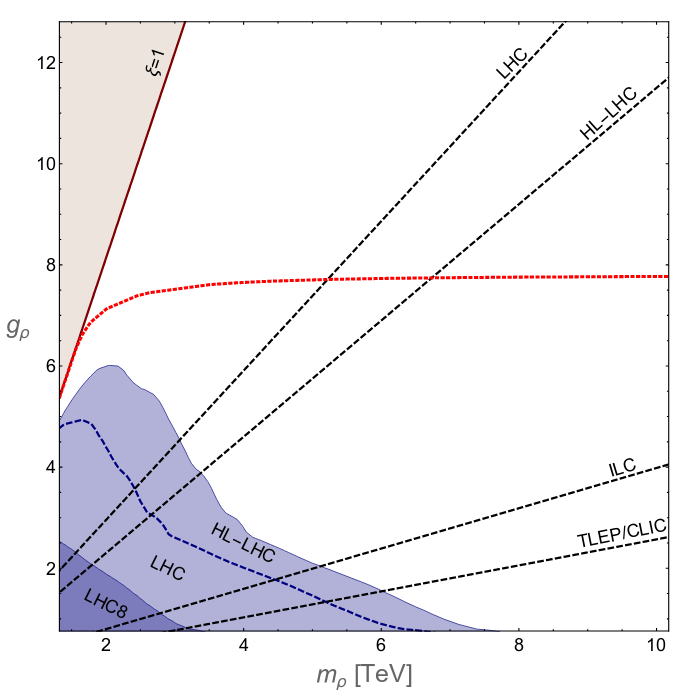}}\\[-1.2cm]
	\includegraphics[height=1.05in]{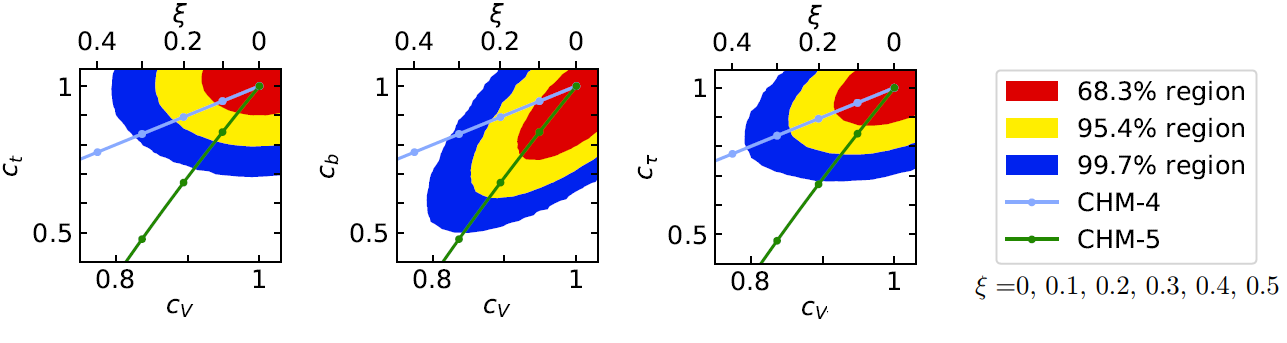}
	\begin{center}	\vspace{-0.6cm}
	\caption{\label{fig:ResHfit} 
	Upper row: Limits from CMS (left) and ATLAS (center) searches for 
	vector resonances \cite{Moriond18} and future projections 
	\cite{Thamm:2015zwa} (right). 
	Lower row:
	Limits on $\xi$ from a fit to Higgs data \cite{deBlas:2018tjm},
	considering the 2D-planes of couplings to gauge bosons and top
	quarks, bottom quarks, and tau leptons, where 
	$c_X = g_{hXX}/g_{hXX}^{\rm SM}$.}
	\end{center}
\end{figure}

An obvious way to test CH scenarios is to search for the composite
resonances predicted in the setup. We first focus on vector-boson
resonances of $Z^\prime$ and $W^\prime$ type, while fermionic resonances
will be discussed further below in Sec.~\ref{sec:tp}.
The most promising decay channels of these resonances in CH models 
clearly involve the heaviest SM states, i.e., $W,Z,h$ bosons or the 
top quark.

From the latest CMS and ATLAS searches, presented at Moriond '18
\cite{Moriond18}, see left plots in Fig.~\ref{fig:ResHfit}, we extract 
the bound
\begin{equation}
m_\rho \gtrsim (2.5-3.5)\,{\rm TeV}\quad {\text{\scriptsize $g_\rho = 3$} \atop \raisebox{4mm}{$\Longleftrightarrow$}} \quad
f \gtrsim (850-1200)\,{\rm GeV}\,,
\end{equation}
which is based on the assumption $g_\rho=3$ and, by now, furnishes
a very competitive constraint.
In fact, the limits depend sensitively on the couplings of the
resonances to the SM fields and have always to be considered with 
care. A thorough analysis, discussing such effects, has been
presented in \cite{Thamm:2015zwa} (see also \cite{Low:2015uha,
Niehoff:2015iaa}), where also projections for future colliders are
obtained. In the rightmost plot in Fig.~\ref{fig:ResHfit}, the $95\%$\,CL
limits from LHC8 as well as projections for the LHC with 300\,fb$^{-1}$
and the high-luminosity LHC with 3000\,fb$^{-1}$ are presented in the $m_\rho - g_\rho$ plane,
employing dark and light violet, respectively.

\subsection{Higgs Physics}

Another powerful test of the CH paradigm is to search for deviations
in the Higgs couplings, potentially unraveling its Goldstone nature.
The couplings to gauge bosons in the MCHM have been given in 
Eq.~\eqref{eq:HVV} while those to fermions are more model dependent. 
For the MCHM$_5$, taking only into account the global shift due to 
the non-linear nature of the Higgs, we obtain (see, e.g., 
\cite{Contino:2013kra})
\begin{equation}
g_{hff} = \frac{1-2\xi}{\sqrt{1-\xi}}\, g_{hff}^{\rm SM}\,.
\end{equation}

Several groups have performed fits to Higgs data, see, e.g.,
\cite{Sanz:2017tco,deBlas:2018tjm,Falkowski:2015fla,
Corbett:2015ksa,Fichet:2015xla,Banerjee:2017wmg,Ellis:2018gqa}.
In the lower panel of Fig.~\ref{fig:ResHfit} we display the results of
Ref.~\cite{deBlas:2018tjm}, which provide limits on the MCHM$_4$
and MCHM$_5$ that can be obtained from the intersection of the 
colored fit regions with the CH predictions, the latter given by blue and 
green lines with dots for various values of $\xi$. The generally 
good agreement with the SM expectations leads to the bound 
\begin{equation}
\xi \lesssim 0.12\ \Leftrightarrow \
f \gtrsim 700\,{\rm GeV} \ \, @ 95\% {\rm CL}
\end{equation}
for the MCHM$_5$ \cite{deBlas:2018tjm,Sanz:2017tco}, which starts 
to become competitive with the one from EWPT.\footnote{For 
current limits in other CH scenarios, see, e.g.,
\cite{Sanz:2017tco,Banerjee:2017wmg}, as well as 
\cite{Bellazzini:2014yua,Azatov:2011qy,Carmona:2013cq,
Barducci:2013wjc,Carena:2014ria}
which include discussions on effects from fermion mixing.}

Note that a detailed comparison of indirect and direct reaches for 
CH models has been performed in \cite{Thamm:2015zwa}, where the direct
limits in the right plot of Fig.~\ref{fig:ResHfit} have been confronted 
with projected indirect limits from Higgs-coupling modifications
at various colliders, given as dashed lines in the same plot.
The latter bound directly the ratio $\xi$ and thus show a linear
behavior in the $m_\rho-g_\rho$ plane (complementary to direct
searches), where projected LHC limits reside in the ballpark of 
$\xi \lesssim 0.1$, while ILC/TLEP/CLIC are expected to reach $\xi
\lesssim 10^{-2}-10^{-3}$.

\section{Light Top Partners and Lepton-Flavor Observables}
\label{sec:tp}

\begin{figure}[!t]
	\begin{center}
	\includegraphics[height=1.55in]{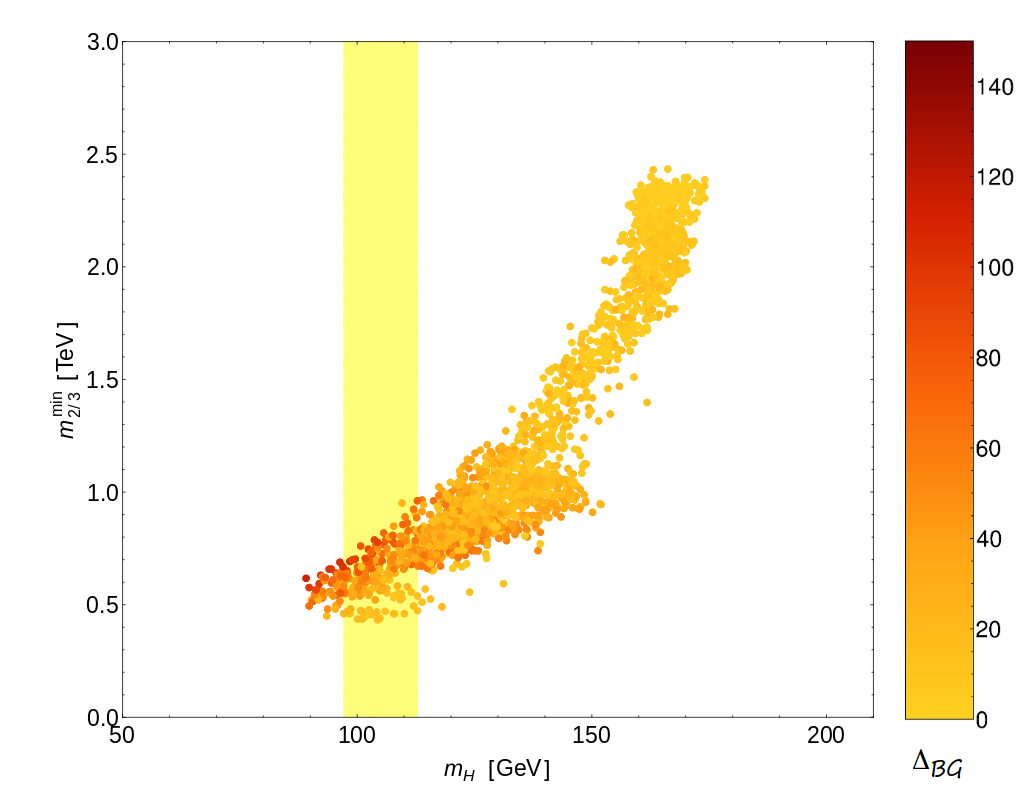}\,
	\includegraphics[height=1.53in]{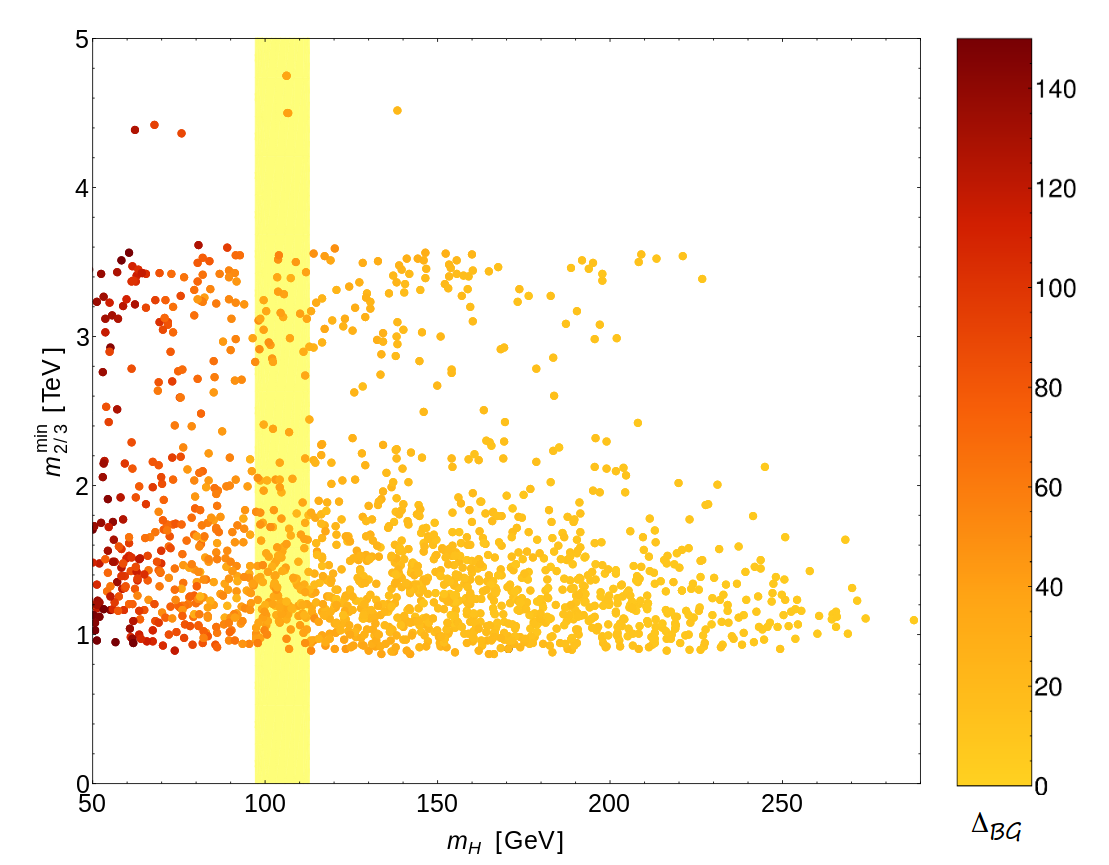}\
	\includegraphics[height=1.64in]{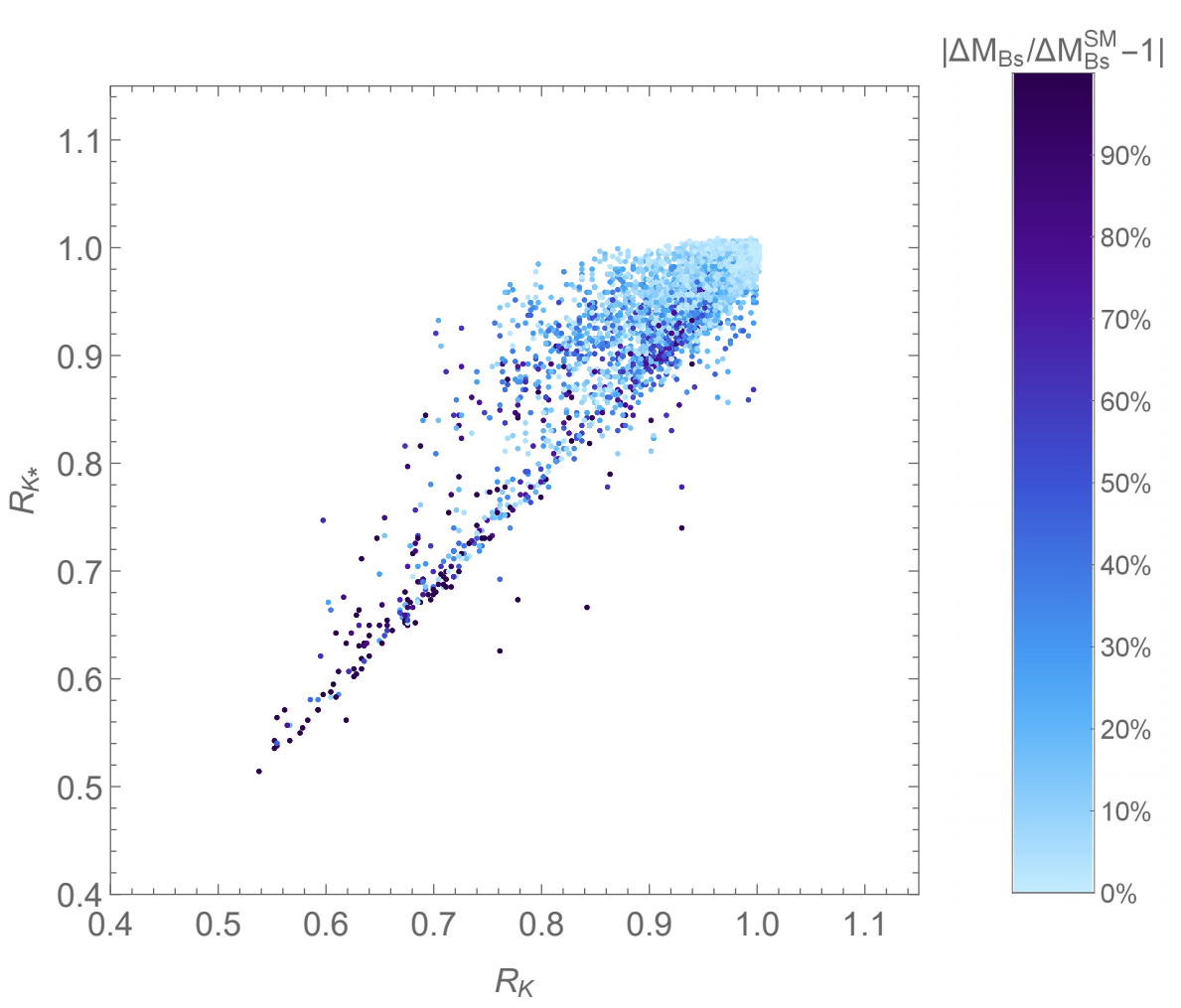}
	\caption{\label{fig:tp} 
	Masses of lightest top partners in the MCHM$_5$ (left) and
	in the minimal lepton model of \cite{Carmona:2015ena}
	(center), where $f=800\,$GeV as well as predictions in the 
	$R_K - R_{K^\ast}$ plane for the latter model.}
	\end{center}
\end{figure}

Before concluding, we finally come back to the issue of light top
partners in CH models. As explained in Sec.~\ref{sec:CH}, we expect
$m^0_T/f \sim m_h/m_t < 1$ in the MCHM$_5$.
The corresponding numerical prediction from \cite{Carmona:2014iwa} is
given in the left plot in Fig.~\ref{fig:tp}, for $f=800\,$GeV, where 
the color code depicts the degree of (Barbieri-Giudice) tuning 
\cite{Barbieri:1987fn}. It turns out that in the 
MCHM$_5$ in fact a 
viable Higgs mass (given by the yellow band) requires generically
\begin{equation}
m^0_T \lesssim 800\,{\rm GeV}\,.
\end{equation}
On the other hand, searches for top partners at ATLAS and CMS are
already excluding masses of up to $\gtrsim 1\,$TeV \cite{Aaboud:2018xuw,
Sirunyan:2017usq}, which is starting to become an issue for TeV-scale 
CH models.

While the masses of top partners could be raised by using less minimal
quark representations, featuring a large number of new states and 
requiring a rather unmotivated ('ad-hoc') tuning in the Higgs mass 
\cite{Panico:2012uw}, an interesting alternative is to consider a
minimal implementation of a non-trivial lepton sector in the MCHM. 
As was shown in \cite{Carmona:2014iwa,Carmona:2015ena}, in the 
framework of the type-III seesaw mechanism it is possible to unify
also the right-handed charged and neutral leptons in a single
representation of the global symmetry, which leads to a highly
predictive lepton sector (featuring 2 instead of 4 $SO(5)$ multiplets)
with in total less degrees of freedom than in standard CH incarnations. 
At the same time, the setup predicts a moderate compositeness of 
the right-handed SM-like charged leptons (driven by the
unification with the seesaw fields) \cite{Carmona:2014iwa}, which 
enters in a parametrically enhanced way into the Higgs potential. This 
non-negligible lepton contribution to the potential allows to raise 
the masses of the top partners, such as to meet  current
experimental limits, while a light Higgs remains natural. 
The numerical results are presented in the central plot of 
Fig.~\ref{fig:tp}, which visualizes that now
the lightest top partners can easily be in the $2-3\,$TeV range.

Another interesting consequence of the scenario of lepton 
compositeness is that it predicts a violation of lepton flavor
universality (LFU), which can be tested for example at LHCb. 
In fact, the latter experiment saw hints for a deviation from 
the SM prediction for the LFU-probing observables
$R_K$ and $R_{K^\ast}$, which reads $R_K^{\rm SM} = 
R_{K^\ast}^{\rm SM} = 1$ to good approximation. It turns out that the
CH model of \cite{Carmona:2015ena} strictly predicts both
$R_K < 1$ and $R_{K^\ast} < 1$, see the right plot in 
Fig.~\ref{fig:tp} \cite{Carmona:2017fsn}, just going in the correct
direction to address the experimental tensions, while meeting 
other constraints from flavor physics \cite{Carmona:2015ena,
Carmona:2017fsn}.

\section{Conclusions}
\label{sec:conc}

We reviewed the Composite Higgs solution to the gauge hierarchy
problem and the flavor puzzle. We collected current constraints
on minimal models, covering EWPT, resonance searches and tests 
of Higgs properties, where the latter two are becoming competitive 
due to the successful LHC operation.
Moreover, we discussed the issue of light top partners in composite
models and presented ways to avoid them with interesting consequences
for flavor physics.

\paragraph{Acknowledgments}
I am grateful to the organizers of ALPS2018 for the invitation
and the excellent organization and atmosphere at the workshop.

\end{document}